\documentclass[twocolumn,showpacs]{revtex4}

\usepackage{bbm,amsmath,amsthm,amssymb,graphicx}

\newcommand{\bra}[1]{{\langle #1|}}
\newcommand{\ket}[1]{{|#1\rangle}}
\newcommand{\proj}[1]{{\ket{#1}\bra{#1}}}

\renewcommand{\sp}[2]{\langle #1|#2\rangle}
\newcommand{\1}{\mathbbm{1}}

\newcommand{\R}{\mathbbm{R}}
\newcommand{\C}{\mathbbm{C}}
\newcommand{\D}{\mathcal{D}}

\newcommand{\B}{\mathcal{B}}

\renewcommand{\S}{\mathcal{S}}

\newcommand{\A}{\mathbbm{A}}
\newcommand{\X}{\mathcal{X}}
\newcommand{\Y}{\mathcal{Y}}
\newcommand{\U}{\mathcal{U}}
\renewcommand{\O}{\mathcal{O}}
\renewcommand{\P}{\mathcal{P}}
\newcommand{\rank}{\mathrm{rank}\,}
\newcommand{\diag}{\mathrm{diag}\,}
\newcommand{\tr}{\mathrm{tr}}
\renewcommand{\H}{\mathcal{H}}
\newcommand{\dm}[1]{\mathcal{S}(#1)}
\renewcommand{\vec}[1]{\mathbf{#1}}
\newcommand{\bl}{{\boldsymbol{\lambda}}}
\newtheorem{lem}{Lemma}

\newtheorem{def2}{Definition}
\renewcommand{\epsilon}{\varepsilon}

\begin{document}

\title{Finite key analysis for symmetric  attacks in quantum key distribution}
\author{Tim Meyer, Hermann Kampermann, Matthias Kleinmann, and Dagmar Bru\ss}
\affiliation{Institut f\"ur Theoretische Physik III, 
Heinrich-Heine-Universit\"at D\"usseldorf, D-40225 D\"usseldorf, Germany}
\begin{abstract}
	We introduce a constructive method to calculate the achievable secret key 
	rate for a generic class of quantum key distribution protocols, when only a
	{\em finite} number $n$ of signals is given. Our approach is applicable to 
	all scenarios in which the quantum state shared by Alice and Bob is known. In 
	particular, we consider the six state protocol with symmetric  eavesdropping 
	attacks, and show that for a small number of signals, i.e. below $n\sim10^4$, 
	the finite key rate differs significantly from the asymptotic value for 
	$n\rightarrow\infty$. However, for larger $n$, a good approximation of the 
	asymptotic value is found. We also study secret key rates for protocols using 
	higher-dimensional quantum systems.
\end{abstract}


\maketitle

\section{Introduction}
The possibility of  secret key distribution is inherent in  quantum mechanics.  
Since the intriguing work of Bennett and Brassard~\cite{ben85:bb84}, who were 
the first to realize this potential, much effort has been devoted to turn their 
idea into feasible protocols for quantum key distribution (QKD).

The aim of a quantum key distribution  protocol is to supply the honest parties 
Alice and Bob with a common, random, and secret bit string. This key is 
generated by Alice sending a number of quantum states to Bob, and Bob   
measuring them randomly in one of a set of bases, previously agreed upon by 
both parties. Equivalently, this process can be seen as the distribution of an
entangled state between Alice and Bob, followed by appropriate measurements on 
both sides~\cite{ben92:ent-based,cur04:ent_as_precondition}. In this paper we 
will use the latter approach, i.e. the entanglement-based formulation. During 
the distribution phase it is unavoidable that the quantum state is disturbed by 
noise, which  -- in the worst case -- has to be attributed to interaction of 
the notorious eavesdropper Eve.

After measuring the shared quantum state, Alice and Bob are left with purely 
classical data, and employ classical algorithms to correct errors and reduce 
the knowledge of Eve. For a given QKD protocol to be unconditionally secure, in 
the end the honest parties must have a perfectly correlated string of bits, 
about which Eve has no knowledge, even though she is given unlimited power 
(i.e. she is only restricted by the laws of physics, but not by any minor 
technological difficulties such as producing a loss-less fiber or building a 
quantum computer). This bit string is the secret key, and its length divided by 
the initial number of signals is the secret key rate. This fundamental quantity 
is, due to the complexity of the various quantum and classical steps, very 
difficult to determine.

Recently, important progress has been achieved towards the calculation of 
secret key rates: unconditional security proofs were formulated for generic QKD 
protocols (see, for 
instance,~\cite{christandl04:generic_sec_proof,ren05:sec_proof_pa}).
In this way, every protocol (e.g. BB84~\cite{ben85:bb84}, B92~\cite{ben92:b92}, 
the Ekert protocol~\cite{eke91:ekert_protocol}, the six-state 
protocol~\cite{bru98:6state,bec98:6state}) can be fit into a common framework 
to analyze the security and derive bounds for the secret key rate. However, 
these bounds only hold for the asymptotic case, where infinitely many signals 
are used. For realistic implementations, it is important to address the case of 
a {\em finite} number of signals. This is the topic of our contribution.

The outline of this article is as follows: in 
section~\ref{sec:h30gfhwsefoj234twef} we give an overview over the structure of
QKD protocols and explain the starting point~\cite{ren05:sec_proof_pa}
of our calculations. In section~\ref{sec:h249ru0jwefksdgv} we review the 
tomographic protocol, before we come to the main part, namely the calculation 
of the entropies for the bound on the secret key rate, in 
section~\ref{sec:8hzg34ghsdi24f2g}. Our results are presented in 
section~\ref{sec:k9gh3hvsbndfvf9234}, and we conclude in 
section~\ref{sec:fh284290uoj4jhre2}.

%%%%%%%%%%%%%%%%%%%%%%%%%%%%%%%%%%%%%%%%%%%%%%%%%%%%%%%%%%%%%%%%%

\section{General quantum key distribution}
\label{sec:h30gfhwsefoj234twef}

In this section we give an overview over the structure of common QKD protocols 
and introduce our notation and some recent results~\cite{ren05:sec_proof_pa}, 
that will be the starting point of our analysis.

Every QKD protocol can be divided into two parts: a quantum part, in which 
quantum mechanical systems are distributed between Alice and Bob and upon which 
some measurements are carried out, yielding classical data. In the second part, 
this data is transformed into a secret key by means of classical  error 
correction and privacy amplification~\cite{maurer93:pa}. We will only consider 
one-way classical post-processing, which will be described in detail below.

\subsubsection{Quantum part}

Most well-known QKD protocols like the BB84~\cite{ben85:bb84}, 
six state~\cite{bru98:6state,bec98:6state}, B92~\cite{ben92:b92},
or the Ekert~\cite{eke91:ekert_protocol}
protocol only differ in the type of  quantum correlations that get distributed
between Alice and Bob, and how much information about the adversary the honest 
parties can extract. The quantum part of the protocol can be summarized by the 
following steps:
\begin{enumerate}
\item[(i)]\emph{Distribution.}
Alice prepares $n'$ maximally entangled states in dimension $d$,
\begin{equation}
	\ket{\phi_d^+}:=\frac{1}{\sqrt d}\sum_{x=0}^{d-1}\ket{xx}\ ,
	\label{eq:bcxvwoe204sdgvs2}
\end{equation}
and sends the second half of each pair to Bob. Due to channel noise and/or
Eve's interference, this state may get corrupted. Thus, after the
distribution, Alice and Bob end up with a state $\rho_{AB}^{n'}$, describing
all $n'$ pairs, that is in general mixed.
\item[(ii)]\emph{Encoding/Measurement.}
Alice and Bob agree on a set of $r$ different encodings (``bases'') 
$\{\ket{e_i^x}\}$ for the qu{\em d}it state $\ket x$, with $1\le i\le r$ and 
$0\le x\le d-1$, where $\sp{e_i^x}{e_i^y}=\delta_{xy}$~\footnote{This can be
generalized even further to include also the case where the quantum  states
encoding the $d$it $x$ are not orthogonal, as it is the case for the
B92 protocol~\cite{ben92:b92}. However, this is not important for our
analysis.}.  For each pair of particles, Alice and Bob choose at random an 
encoding $j$ and $k$ and measure their particles with respect to that basis.
They obtain a classical $d$it value, where the correlation between these $d$its 
depends on the choice of the encodings. As an example, in the two-dimensional 
case ($d=2$) for the BB84 protocol, we have $r=2$ different encodings: 
$\ket{e_{1,2}^x}$, with $x=0,1$, are the eigenstates of two Pauli operators.
\item[(iii)]\emph{Parameter estimation.}
By comparing a random part of the data collected during the measurement step, 
Alice and Bob can get some information about the state $\rho_{AB}^{n'}$.  
Usually, this will be the error rate, which can be calculated for all different 
encodings used in the previous step. Depending on this information, Alice and 
Bob decide whether to continue with the protocol or abort, if they cannot 
ensure its security.
\item[(iv)]\emph{Sifting.}
Alice and Bob announce over the classical channel which encoding they chose for 
each qu$d$it pair. All their measurement data for which the setting 
matched~\footnote{There exist more sophisticated sifting strategies, e.g. in 
the SARG protocol~\cite{sca04:sarg}.} form the sifted keys $\vec X$ and $\vec 
Y$ for Alice and Bob, respectively, which are not necessarily identical yet. We 
denote by $n$ the length of the strings $\vec X$ and $\vec Y$ \emph{after} the 
sifting step, i.e. the number of states that were measured in the same basis by 
Alice and Bob. This means that $n$ is approximately equal to $n'$ divided by 
the number of different encodings used in step (ii). We denote by $\rho_{AB}^n$ 
the part of the state $\rho_{AB}^{n'}$ which is kept in the sifting.
\end{enumerate}
At this point, Alice and Bob are left with purely classical data, namely the 
$d$it strings $\vec x$ and $\vec y$, whereas Eve might still hold a quantum 
system that was entangled with $\rho_{AB}^n$. We have to consider the worst 
case, in which Eve holds a purifying system of $\rho_{AB}^n$, i.e.  
$\rho_{AB}^n=\tr_E\proj{\psi_{ABE}}$, where the system in $E$ is under Eve's 
control. The situation where classical data (which is obtained from 
$\rho_{AB}^n$ by Alice's and Bob's measurements) is correlated with a quantum 
system can be described by a classical-classical-quantum  state~\cite{dev03:cq}
\begin{equation}
	\rho_{\vec X\vec YE}=\sum\limits_{\vec x,\vec y}P_{\vec X\vec Y}
	(\vec x,\vec y)P_\ket{\vec x}\otimes P_\ket{\vec y}\otimes
	\rho_E^{\vec x\vec y}.
	\label{eq:bvbw9238vsd0gvwka}
\end{equation}
Here, $P_{\vec X\vec Y}$ is the probability distribution of Alice's and Bob's 
random variables $\vec X$ and $\vec Y$ and $\rho_E^{\vec x\vec y}$ is the state 
that Eve holds if $\vec X=\vec x$ and $\vec Y=\vec y$. We use the notation that 
capital letters, e.g. $X$, represent classical random variables, taking values 
$x$ from an alphabet $\X=\{0,1,\dots,d-1\}$. Bold letters denote vectors, e.g.  
$\vec x=(x_1,x_2,\dots,x_n)$. We denote by $P_{\ket x}=\proj x$ the projector 
on the quantum state $\ket x$ and by $P_X$ the probability distribution of the 
random variable $X$.

\subsubsection{Classical part}

In this part of the key distribution, which is common for all well-known QKD 
protocols (with one-way post-processing), the classical strings $\vec X$ and 
$\vec Y$ will be made equal and secure. This is achieved by the following
classical sub-protocols:

\begin{enumerate}
\item[(v)]\emph{Pre-processing and error correction.}
In the pre-processing stage Alice computes a new random variable $\vec U$ from 
her data $\vec X$ by the use of the channel $\vec U\leftarrow\vec X$, defined 
by some conditional  probability distribution $P_{\vec U|\vec X}$. The string 
$\vec U$ will  then serve as the key. In the error correction step, Alice sends 
the information that Bob needs to compute $\vec U$ from his data $\vec Y$. This 
information can be quantified by a random variable $\vec W$.
\item[(vi)]\emph{Privacy amplification.}
Alice and Bob shrink the length of the key $\vec U$ and at the same time  
reduce the information that Eve might have about it, thereby generating
a secret key. Since the privacy amplification is an important step, which will 
be the starting point of our calculation, we review this sub-protocol in more 
detail here. We also review the security analysis of privacy amplification and 
present an expression for an achievable secret key length, as found 
in~\cite{ren05:pa}.
\end{enumerate}

\subsubsection*{Secret key generation by privacy amplification}

Consider the case in which Alice and Bob hold a common random string $\vec U$, 
which is supposed to serve as a secret key. In the privacy amplification step, 
the information that Eve might have about the key $\vec U$ is reduced. This is 
done by choosing a two-universal hash function $F$ and computing $F(\vec U)$ as 
the new key. A two-universal hash function is a random function
$F:\U\rightarrow\{0,1\}^\ell$ such that $F(u)$ and $F(u')$ are independent and 
uniformly distributed for all $u\ne u'$~\cite{ren05:pa}. Then the information 
that Eve can have about $F(\vec U)$, depending on the quantum state $\rho_E$ 
she holds, can be bounded~\cite{ren05:pa}. This result can be applied to 
calculate the secret key rate obtainable by Alice and Bob. ``Secrecy'' is 
measured with respect to the universal composable definition of unconditional 
security~\cite{ben05:universal_composable}: Let $\vec S_A$ and $\vec S_B$ be 
random variables that describe keys that Alice computes from $\vec U$ and Bob 
computes from his guess about $\vec U$, using the random hashing. This 
situation, together with Eve holding a quantum state containing some 
information about the keys, can be described by the classical-classical-quantum 
state $\rho_{\vec S_A\vec S_B E}$. The case of a perfect key, i.e. $\vec 
S_A=\vec S_B=\vec S$, where $\vec S$ is uniformly distributed over the set of 
all possible keys $\S$ and Eve being completely uncorrelated with $\vec S$ is 
described by $\rho_{\vec S\vec S}\otimes\rho_E:=1/|\S|\sum_{\vec 
s\in\vec\S}P_\ket{\vec s}\otimes P_\ket{\vec s}\otimes \rho_E$. The key pair 
$\vec S_A,\vec S_B$ is said to be $\epsilon$-secure, if $\|\rho_{\vec S_A\vec 
S_B E}-\rho_{\vec S}\otimes\rho_E\|\le \epsilon$. Here, 
$\|\rho-\sigma\|=\tr|\rho-\sigma|/2$, with $|A|=\sqrt{A^\dagger A}$, denotes 
the trace distance between $\rho$ and $\sigma$. It provides a measure of how 
close the actual system is to the ideal case and how ``secure'' the final key 
will be. 

An important result which will be used here was found in~\cite{ren05:pa} (see 
also~\cite{ren05:sec_proof_pa}): Suppose Alice and Bob both share the same 
random string $\vec U$, which they compute from their raw data strings
$\vec X$ and $\vec Y$ via pre-processing and error correction. The adversary 
holds some quantum system $\rho_E$ that might be correlated with $\vec U$, i.e, 
the total system can be represented by some density operator $\rho_{\vec UE}$.  
Then an achievable length $\ell$ of the secret key that can be computed from 
$\vec U$ by a two-universal hash function $F$ is given by~\cite{ren05:pa}:
\begin{equation}
	\ell =S_2^{\epsilon'}(\rho_{\vec UE})-S_0^{\epsilon'}(\rho_E)
	-2\log_2(1/\epsilon),
	\label{eq:grehza9sd8f24}
\end{equation}
with $\epsilon'=(\epsilon/8)^2$, if the key is required to be $\epsilon$-secure 
with respect to $\rho_E\otimes P_{\ket F}$. Here, the state $\rho_E\otimes 
P_{\ket F}$ describes the total knowledge of Eve, since she also learns the 
function $F$ on which Alice and Bob have to agree by public communication. The 
quantities $S_2^\epsilon$ and $S_0^\epsilon$ that occur in 
Eq.~(\ref{eq:grehza9sd8f24}) are called \emph{smooth Renyi entropies} and are 
defined as follows.

Denote by $\B^\epsilon(\rho)$ the set of density matrices that are 
$\epsilon$-close to $\rho$, i.e.  
$\B^\epsilon(\rho):=\{\sigma\in\dm{\H}:\|\rho-\sigma\|\le\epsilon\}$, where 
$\dm{\H}$ is the set of density matrices acting on the Hilbert space $\H$.
\begin{def2}
	\label{def:entropy_quantum}
	Let $\rho\in\dm\H$ and $\epsilon\ge0$.
	The $\epsilon$-smooth Renyi entropies of order 2 and 0 are defined as
	\begin{eqnarray}
		S_2^\epsilon(\rho)&=&-\log_2\inf_{\sigma\in\B^\epsilon(\rho)}\tr\sigma^2,\\
		S_0^\epsilon(\rho)&=&\log_2\inf_{\sigma\in\B^\epsilon(\rho)}\rank\sigma.
	\end{eqnarray}
\end{def2}
In the following, we will also need a classical Renyi entropy, which is defined 
as follows:
\begin{def2}
	Let $X$ and $Y$ be random variables, taking values $x\in\X$ and $y\in\Y$,
	respectively, and let $P_{XY}$ be their probability distribution. Then the 
	conditional smooth Renyi entropy of order zero is defined 
	as~\cite{ren05:renyi}
	\begin{eqnarray}
		&&H_0^\epsilon(X|Y):=\min\limits_{\A:P(\A)\ge1-\epsilon}\hspace{4cm}\nonumber\\
		&&\hspace{1cm}\left(\max\limits_{y\in\Y}\log_2|\{x\in\X:P_{X\A|Y=y}(x)>0\}|
		\right).
	\end{eqnarray}
\end{def2}
Here, the minimum is taken over all events $\A$ that occur with probability at 
least $1-\epsilon$. Smooth Renyi entropies are generalizations of the 
conventional Renyi entropies~\cite{renyi60:entropy}: The classical Renyi 
entropy of a probability distribution $P_X$ is a measure of the largest (in the 
case of $S_2$) or smallest (in the case of $S_0$) uncertainty about $X$ that 
can be found within all probability distributions that are close~\footnote{The 
distance between two classical probability distributions $P_X$ and $Q_X$ is 
measured by the variational distance $\|P-Q\|=1/2\sum_{x\in\X}|P(x)-Q(x)|$.} to 
$P_X$. In the quantum case, this translates to the entropy of density operators 
that have a trace distance to $\rho$ that is less or equal to $\epsilon$.

If we want to apply Eq.~(\ref{eq:grehza9sd8f24}) to our QKD protocol, we 
need to specify the overall quantum state representing Alice's and Bob's
classical strings and the information that Eve holds, which might
be at least partly of quantum nature: It consists of a density
operator $\rho_E^{\vec x\vec y}$ that depends on the strings $\vec x$
and $\vec y$ that Alice and Bob have measured, together with the
classical information that is interchanged via the public channel,
i.e. the error correction information $\vec w$. After the error
correction, Alice and Bob both hold the same string $\vec u$.
Thus, the situation can be described by the following quantum state:
\begin{equation}
	\rho_{\vec U\vec WE}=\sum\limits_{\vec x,\vec y,\vec u,\vec w}
	P_{\vec X\vec Y\vec U\vec W}(\vec x,\vec y,\vec u,\vec w)
	P_{\ket{\vec u}}\otimes(P_{\ket{\vec w}}\otimes\rho_E^{\vec x\vec 
	y}).
\end{equation}
If we now use Eq.~(\ref{eq:grehza9sd8f24}) to calculate the key
length, we still have the dependence on the error correction
information $\vec W$. In~\cite{ren05:sec_proof_pa} it was shown
that it can be removed, leading to another additive term
$H_0^\epsilon(\vec U|\vec Y)$, which is the information needed to correctly 
guess $\vec U$ from $\vec Y$ with probability of at least $1-\epsilon$. 
The quantity $H_0^\epsilon$
is called (classical) \emph{conditional smooth Renyi entropy}, and was
defined
above. We will restrict ourselves to the
simple case where Alice skips the pre-processing step (first part of step
(v) in our generic protocol, cf. section~\ref{sec:h30gfhwsefoj234twef}), 
i.e. $\vec U=\vec X$. This leads to the following formula for an achievable 
length of the $\epsilon$-secure key, which 
will be the starting point of our calculations:
\begin{equation}
	\ell =S_2^{\epsilon'}(\rho_{\vec XE})-S_0^{\epsilon'}(\rho_E)
	-H_0^{\epsilon'}(\vec X|\vec Y)-2\log_2(1/\epsilon),
	\label{eq:jkfhw38hfa0gv3w4t}
\end{equation}
with $\epsilon'=(\epsilon/8)^2$. 

\section{The tomographic protocol}
\label{sec:h249ru0jwefksdgv}

Although equation~(\ref{eq:jkfhw38hfa0gv3w4t}) is an explicit formula
for an achievable key length for any QKD protocol that fits into the 
framework described
in section~\ref{sec:h30gfhwsefoj234twef}, the main problem is the
ignorance about Eve's state $\rho_E$. If this state is not known, the
entropies in Eq.~(\ref{eq:jkfhw38hfa0gv3w4t}) cannot be calculated. However,
the data gathered in the parameter estimation step (iii) poses 
some restrictions on Eve's state. For example,
in the BB84 protocol, starting from $\ket{\phi^+}$ as defined 
in~(\ref{eq:bcxvwoe204sdgvs2}) with $d=2$,
a measured bit error rate $e_b$ implies a fraction
$e_b$ of $\proj{\psi^+}$ and $\proj{\psi^-}$ of the $n$ qubits
shared by Alice and Bob (here $\ket{\phi^\pm}$ and $\ket{\psi^\pm}$ are
the usual Bell states). Thus it is possible to deduce part of the
structure of Eve's purification. Exploiting this knowledge, one can obtain
a lower bound on Eq.~(\ref{eq:jkfhw38hfa0gv3w4t}) by taking the infimum 
over all states of Eve that are compatible with the statistics obtained
in the parameter estimation step. Having this in mind, we make
the following assumptions for our finite key analysis:
\begin{enumerate}
\item\emph{Collective attack.} 
The state that Alice and Bob share after
the distribution step is given by
\begin{equation}
	\rho_{AB}^{n'}=\rho_{AB}^{\otimes n'},
	\label{eq:jgwri0g2sd0gj24gwe4}
\end{equation}
i.e. Eve interacts only with individual signals and does so in the same
way for all copies. Note that this is not really a restriction, since 
in~\cite{ren05:sec_proof_pa} it was shown that Alice and Bob can always 
symmetrize the state $\rho_{AB}^{n'}$ to a tensor product form by only slightly 
modifying the protocol.
\item\emph{Symmetric attack.}
Each single state is a depolarized version of the maximally entangled
state~(\ref{eq:bcxvwoe204sdgvs2}), i.e.
\begin{equation}
	\rho_{AB}=(\beta_0-\beta_1)\proj{\phi_d^+}+\frac{\beta_1}{d}\1.
	\label{eq:n2492mvsssdfsasaa}
\end{equation}
\end{enumerate}
We have adopted here the notation of~\cite{bruss03:tomographic_qkd}. The two 
parameters $\beta_0$ and $\beta_1$ are not independent, and the normalization 
condition reads $\beta_0+(d-1)\beta_1=1$. One can interpret $\beta_0$ as the 
probability that Alice and Bob get the same output, and $\beta_1$ as the 
probability that they get a particular other one, so we always assume 
$0\le\beta_1<1/d<\beta_0\le1$. In the limit $n\rightarrow\infty$, the error 
rate in the sifted key (for $d=2$, this is called the \emph{quantum bit error 
rate}, QBER) is given by $1-\beta_0=(d-1)\beta_1$.

We make no further restrictions on the eavesdropping strategy besides being 
collective and symmetric, and therefore assume that Eve holds the purifying 
system of each state $\rho_{AB}$.  It is important to note that by fixing the 
form of the distributed state $\rho_{AB}^n$ (as in 
Eq.~(\ref{eq:jgwri0g2sd0gj24gwe4}) and (\ref{eq:n2492mvsssdfsasaa})),
which is the ``output'' of the whole quantum part of the protocol (cf.  
section~\ref{sec:h30gfhwsefoj234twef}), the encoding step (ii) essentially 
becomes meaningless. This is because we now have the freedom to choose any kind 
of encoding, since in the end we are assuming $\rho_{AB}^n=\rho_{AB}^{\otimes 
n}$ with $\rho_{AB}$ given by (\ref{eq:n2492mvsssdfsasaa}) anyway. However, as 
Eve knows Alice's and Bob's protocol, she would not necessarily conduct such a 
symmetric attack if Alice and Bob could not check for the state $\rho_{AB}$ to 
be of the form (\ref{eq:n2492mvsssdfsasaa}). Therefore, we assume that Alice 
and Bob use a scheme that enables them to do so, which is achieved by encoding 
the basis states $\{\ket x\}$ into $d+1$ mutually unbiased bases, which 
corresponds to a generalization of the six-state protocol to $d$ dimensions 
(where $d$ is a prime power). Such a ``tomographic'' protocol was originally 
suggested
in~\cite{bruss03:tomographic_qkd,liang03:tqc}, in the context of a connection 
between advantage distillation and entanglement distillation. In the parameter 
estimation step (iii), the only unknown parameter $\beta_0$ (or $\beta_1$) in 
Eq.~(\ref{eq:n2492mvsssdfsasaa}) can be estimated by comparing a randomly 
chosen subset of the raw key.  

%%%%%%%%%%%%%%%%%%%%%%%%%%%%%%%%%%%%%%%%%%%%%%%%%%%%%%%%%%%%%%%%%

\section{Method for calculating smooth Renyi entropies}
\label{sec:8hzg34ghsdi24f2g}

In this section, we derive a method for calculating an achievable secret key 
length for our generic protocol introduced in the previous section. This method 
is applicable in all scenarios, in which the state $\rho_{AB}^n$ is known.  
Explicitly, we will study as an example the state for a symmetric attack, as 
defined via Eqns.~(\ref{eq:jgwri0g2sd0gj24gwe4}) and 
(\ref{eq:n2492mvsssdfsasaa}). For a given state $\rho_{AB}^n$ (and its 
purification), the difficulty in computing the key 
length~(\ref{eq:jkfhw38hfa0gv3w4t}) is due to the minimization over the 
$\epsilon$-environment $\B^\epsilon(\rho)$ involved in the (quantum) smooth 
Renyi entropies. This is because very little is known about the structure of 
the set of density matrices that are close to a given one.  Even for states 
with a tensor structure, as for $\rho_{AB}^{\otimes n}$ in our case, the 
analysis is still very involved, since $\B^\epsilon(\rho^{\otimes n})$ of 
course not only contains product states.  Fortunately, since we are only 
interested in minimizing a function of
the eigenvalues of density matrices, it turns out that we can restrict our 
attention to matrices which have the same eigenvectors as $\rho$. This 
intuition is formalized in Lemma~\ref{lem:hg42810psafh2rfhj}.

Let us denote by $\bl(\rho)$ the ordered spectrum of $\rho$, i.e.
$\bl(\rho)=(\lambda_1,\dots,\lambda_d)\in\R^d$ in ascending order, with 
$d=\dim(\H)$. Also denote by $\|\bl-\bl'\|=1/2\sum_i|\lambda_i-\lambda'_i|$ the 
distance of the vectors $\bl$ and $\bl'$. Recall that $\B^\epsilon(\rho)$ is 
the set of density matrices which are $\epsilon$-close to $\rho$. We define 
$\D^\epsilon(\rho)=\{\sigma\in\dm\H:[\sigma,\rho]=0,\|\bl(\sigma)-\bl(\rho)\|
\le\epsilon\}$ to be the set of density matrices which commute with $\rho$ 
(i.e. they have the same eigenvectors) and have a spectrum $\epsilon$-close to 
that of $\rho$.

\begin{lem}
	The two sets $\Lambda_\B^\epsilon(\rho)=\{\bl(\sigma):\sigma\in
	\B^\epsilon(\rho)\}$, and $\Lambda_\D^\epsilon(\rho)=\{\bl(\sigma):\sigma\in
	\D^\epsilon(\rho)\}$, defined as the sets
  of spectra that correspond to the sets of density matrices
	$\B^\epsilon(\rho)$ and $\D^\epsilon(\rho)$, respectively, are identical.
	\label{lem:hg42810psafh2rfhj}
\end{lem}
\begin{proof}
	Since for two commuting matrices $\rho$ and $\sigma$, we have that
	$\|\rho-\sigma\|=\|\bl(\rho)-\bl(\sigma)\|$, it follows immediately that
	$\D^\epsilon(\rho)\subset\B^\epsilon(\rho)$ which in turn implies
	$\Lambda_\D^\epsilon(\rho)\subset\Lambda_\B^\epsilon(\rho)$.
	The other inclusion follows from the fact~\cite{fan73:trace_distace} that
	$\|\rho-\sigma\|\ge\|\bl(\rho)-\bl(\sigma)\|$.
\end{proof}
From this lemma, it follows immediately that all functions than only depend on 
the eigenvalues of a density matrix and which are to be minimized over the set 
$\B^\epsilon(\rho)$ can equivalently be minimized over $\D^\epsilon(\rho)$.  In 
particular, this holds for the smooth Renyi entropies defined in 
Def.~\ref{def:entropy_quantum}.

Our goal is to calculate the achievable key rate in the case where Alice and 
Bob hold an $n$-fold tensor product of the state $\rho_{AB}$, as defined by 
Eq.~(\ref{eq:n2492mvsssdfsasaa}). It was shown in~\cite{liang03:tqc} that a 
purification of the state~(\ref{eq:n2492mvsssdfsasaa}) is given by
\begin{equation}
	\ket\Psi=\sqrt\frac{\beta_0}{d}\sum\limits_{k=0}^{d-1}\ket{kk}\ket{E_{kk}}+
	\sqrt\frac{\beta_1}{d}\sum\limits_{k\ne l}\ket{kl}\ket{E_{kl}},
	\label{eq:dfje82wjl4fr434tos}
\end{equation}
where Eve's states $\ket{E_{kl}}$ are constrained by 
$\sp{E_{kk}}{E_{ll}}=1-\beta_1/\beta_0$ for $k\ne l$ and $\ket{E_{kl}}$ is 
orthogonal to all other states for $k\ne l$.

To calculate the key length~(\ref{eq:jkfhw38hfa0gv3w4t}), we need to know the 
states $\rho_{\vec XE}=1/d^n\sum_{\vec x,\vec y}P_{\vec X\vec Y}(\vec x,\vec y)
P_{\ket{\vec x}}\otimes\rho_E^{\vec x\vec y}$ and $\rho_E=\tr_{\vec 
X}\rho_{\vec XE}$. Bob's random variable $\vec Y$ does not appear here 
explicitly, since it is equal to that of Alice after the error correction.
From Eq.~(\ref{eq:dfje82wjl4fr434tos}), we can readily compute $\rho_E^{\vec 
x\vec y}$, which is the state that Eve holds if Alice and Bob got the string 
$\vec x$ and $\vec y$ as their measurement results, as well as the 
probabilities $P_{\vec X\vec Y}(\vec x,\vec y)$. We find
\begin{eqnarray}
	\rho_{\vec XE}&=&\left[\frac{1}{d}\sum\limits_xP_{\ket x}\otimes
	\left(\beta_0P_{\ket{E_{xx}}}+\beta_1\sum\limits_{
	\genfrac{}{}{0pt}{}{y}{y\ne x}}
	P_{\ket{E_{xy}}}\right)\right]^{\otimes n}
	\label{eq:fj240sdifh02jfslaf}        \\
	\rho_E&=&\left[\frac{1}{d}\left(\beta_0\sum\limits_{x}P_{\ket{E_{xx}}}
	+\beta_1\sum\limits_{\genfrac{}{}{0pt}{}{x,y}{y\ne x}}P_{\ket{E_{xy}}}
	\right)\right]^{\otimes n}
	\label{eq:dnfjib7qwg92sdg4}
\end{eqnarray}
\begin{equation}
	P_{\vec X\vec Y}(\vec x,\vec y)=\prod\limits_{i=1}^n\left[\beta_1+
	\delta_{x_iy_i}(\beta_0-\beta_1)\right].
	\label{eq:h024hsdbaejfd234}
\end{equation}
Due to the properties of the states $\ket{E_{xy}}$, it follows that Eve's
state
$\rho_E$ has rank  $d^{2n}$ if $\beta_0\ne1$, and
$\rank\rho_{\vec XE}=d^{3n}$ if $\beta_0\ne1$.

In the following three subsections, we analytically compute the entropies
that appear in Eq.~(\ref{eq:jkfhw38hfa0gv3w4t}). It turns out that they
are given by simple functions of the eigenvalues of the corresponding
density matrices. Unfortunately, they
cannot be expressed in a closed form, so in the end we have to resort to
numerics in order to obtain numbers. However, all numerical calculations
stay exact without any approximations, and can be performed in a very
efficient way. We explain the calculations of the entropies in
 some detail, since we believe that our method is interesting and 
useful on its own, as it can be used whenever one wants to determine
the extremum for a function of the spectrum of a state,
in the neighborhood of a given density matrix.

\subsection{Calculation of $S_0^\epsilon(\rho_E)$}

To calculate $S_0^\epsilon(\rho_E)$, we need the eigenvalues of the state 
$\rho_E\in\dm{(\C^d)^{2n}}$. It turns out that $\rho_E$, as defined in 
Eq.~(\ref{eq:dnfjib7qwg92sdg4}), has the following eigenvalues $\lambda_l$ and 
corresponding multiplicities $n_l$, for $0\le l\le n$, where $n$ is the number 
of signals after the sifting:
\begin{eqnarray}
	\lambda_l&:=&\left(\beta_0-\beta_1+\frac{\beta_1}{d}\right)^l
	\left(\frac{\beta_1}{d}\right)^{n-l}\\
	&=&\left(\frac{\beta_0(d+1)-1}{d}
	\right)^l\left(\frac{1-\beta_0}{d(d-1)}\right)^{n-l}\\
	n_l&:=&\genfrac(){0pt}{}{n}{l}(d^2-1)^{n-l}
\end{eqnarray}
Note that the $\lambda_l$ are given in ascending order. We will use
the convention that $\lambda_l$ denotes all \emph{different}
eigenvalues of $\rho_E$, and therefore an index of $\lambda$ runs 
from 0 to $n$, although there are $d^{2n}$ eigenvalues in total,
which we will denote by $\lambda'$, such that 
$\{\lambda_l\}_{0\le l\le n}=\{\lambda'_{l'}\}_{1\le l'\le d^{2n}}$.

Now and in the following we use Lemma~\ref{lem:hg42810psafh2rfhj},
which allows us to calculate the infimum in 
$S_0^\epsilon(\rho_E)=\log_2\inf_{\sigma\in\B^\epsilon(\rho)}\rank\sigma$ 
by only varying the eigenvalues of $\rho_E$. Thus we are looking for
a density matrix $\sigma$ with eigenvalues $\{\mu_i\}$ which is diagonal
in the same basis as $\rho_E$ and has rank
as small as possible under the constraints 
$\sum_i|\lambda'_i-\mu_i|\le2\epsilon$.  Clearly, such a matrix is given
by
$\sigma=\diag(0,\dots,0,\lambda'_{k+1},\dots,\lambda'_{d^{2n}-1},
\lambda'_{d^{2n}}+\delta)$, where 
$\sum_{i=1}^k\lambda'_i=:\delta\le\epsilon$ with $k$ chosen maximally. 
In this way we have found $\rank\sigma=\rank\rho_E-k$.
It remains to determine $k$, which can be done efficiently because of 
the degeneracy of the eigenvalues. Below we construct an algorithm
that computes $k$ in $\O(n)$ running time, rather than scaling with
the total number of eigenvalues $\O(d^n)$.

In order to calculate $k$, define 
\begin{equation}
	s_r:=\sum\limits_{i=1}^rn_{i-1}\lambda_{i-1},
\end{equation}
for $\quad0\le r\le n+1$, which is the sum of the $r$ smallest \emph{different} 
eigenvalues. (For $r=0$, the sum is taken to be zero.) Moreover, let
\begin{equation}
	b:=\max\{r:s_r\le\epsilon\}
\end{equation}
be the the largest number such that the sum of the $b$ smallest 
\emph{different} eigenvalues is smaller than $\epsilon$. A moment of thinking 
then reveals that $k$ is given by
\begin{equation}
	k=\sum\limits_{i=1}^bn_{i-1}+\left\lfloor\frac{\epsilon-s_b}{\lambda_b}
	\right\rfloor,
\end{equation}
%To also cover the (unreasonable) case where $\epsilon\ge1$ we may set
%$k=d^{2n}-1$ if $b=n+1$, which is the maximal number of eigenvalues that can
%be put to 0, leading to $\sigma=\diag\{0,\dots,0,1\}$.
%Having calculated $k$, the entropy is finally given by
where $\lfloor x\rfloor$ denotes the largest integer smaller than or equal to 
$x$. This leads to
\begin{equation}
	S_0^\epsilon(\rho_E)=\log_2(d^{2n}-k).
\end{equation}

\subsection{Calculation of $S_2^\epsilon(\rho_{\vec XE})$}
\label{ss:s2}

The calculation of $S_2^\epsilon(\rho_{\vec XE})$ is similar to the calculation 
of $S_0^\epsilon(\rho_E)$. We first need the eigenvalues of $\rho_{\vec 
XE}\in\dm{(\C^d)^{3n}}$, defined in Eq.~(\ref{eq:fj240sdifh02jfslaf}), and 
their multiplicities. This matrix has $d^{2n}$ non-zero eigenvalues in total:
\begin{eqnarray}
	\lambda_{l+1}&:=&\left(\frac{\beta_0}{d}\right)^l
	\left(\frac{\beta_1}{d}\right)^{n-l}\\
	&=&\frac{1}{d^n}\beta_0^l
	\left(\frac{1-\beta_0}{d-1}\right)^{n-l}\\
	n_{l+1}&:=&d^n\genfrac(){0pt}{}{n}{l}(d-1)^{n-l},
\end{eqnarray}
for $0\le l\le n$. Moreover, $\rho_{\vec XE}$ has $d^{3n}-d^{2n}$ zero 
eigenvalues, independently of $\beta_0$ and $\beta_1$:
\begin{eqnarray}
	\lambda_0&:=&0\\
	n_0&:=&d^{3n}-d^{2n}
\end{eqnarray}
Altogether, we have $0\le l\le m$, with $m:=n+1$, denoting all different 
eigenvalues.

\begin{figure}[tbp]
	\begin{centering}
		\includegraphics[width=8cm]{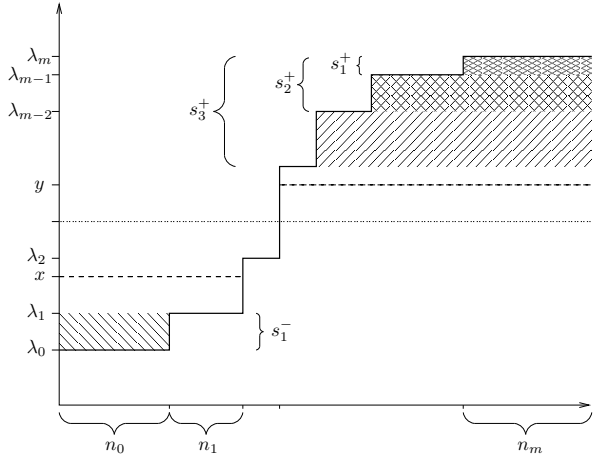}
		\caption{Visualization of the definition of $x$, $y$, and $s_r^\pm$,
		together with the eigenvalues $\lambda_l$ and multiplicities $n_l$
		as defined in section~\ref{ss:s2}. In this example, we have $b^-=1$
		and $b^+=3$.}
	\label{fig:xcvhsid2h0shdfgg}
	\end{centering}
\end{figure}

Recall that 
$S_2^\epsilon(\rho)=-\log_2\inf_{\sigma\in\B^\epsilon(\rho)}\tr\sigma^2$,
thus we are looking for a density matrix $\sigma$ with ordered 
eigenvalues $\{\mu_i\}$ that minimizes
$\sum_i\mu_i^2$ under the constraints $\sum_i\mu_i=1$ and 
$\sum_i|\lambda_i-\mu_i|=2\epsilon$.  Using the Lagrange multiplier
method it can be shown that the solution is
\begin{equation}
	\mu_i=\left\{\begin{array}{ccr} x & \mbox{for} & 0\le i\le b^- \\
	\lambda_i & \mbox{for} & b^-<i<n-b^+ \\
	y & \mbox{for} & n-b^+\le i\le m\end{array}\right.,
\end{equation}
with some constants $x,y,b^-,b^+$ which have to be determined. 
This means that the smallest $b^-+1$
eigenvalues $\lambda$ get raised to $x$, the largest $b^++1$ get lowered to 
$y$, and the intermediate ones stay unchanged. Since the mean 
$\sum_i\mu_i/d^{3n}$ has to remain $1/d^{3n}$, we find $y$ and $x$ by
cutting the largest (smallest) eigenvalues such that the sum of
differences between the largest (smallest) ones and $y$ ($x$) equals $\epsilon$
(see also Fig.~\ref{fig:xcvhsid2h0shdfgg}).

In the following, we give an efficient algorithm for calculating the 
constants $x,y,b^-$, and $b^+$, which is very similar to the one calculating
$S_0^\epsilon(\rho_E)$ in the previous section. Let
\begin{eqnarray}
	s_r^+&:=&\sum\limits_{i=1}^rn_{m-i+1}(\lambda_{m-i+1}-
	\lambda_{m-r}),\\
	s_r^-&:=&\sum\limits_{i=1}^rn_{i-1}(\lambda_r-\lambda_{i-1}),
\end{eqnarray}
for $0\le r\le m=n+1$.
Then the number of the largest (smallest) \emph{different} eigenvalues, 
that can be lowered to $y$ (raised to $x$) is given by
\begin{equation}
	b^\pm:=\max\{r:s_r^\pm\le\epsilon\}.
\end{equation}
With these definitions we find that
\begin{eqnarray}
	x&=&\lambda_{b^-}+\frac{\epsilon-s_{b^-}}{\sum_{i=0}^{b^-}n_i},\\
	y&=&\lambda_{m-b^+}-\frac{\epsilon-s_{b^+}}{\sum_{i=0}^{b^+}n_{m-i}}.
\end{eqnarray}
%A special case occurs if $\epsilon$ is so large that $x$ becomes
%larger and/or $y$ becomes smaller than the average $1/d^{3n}$.
%In that case the solution is simply $\mu_i=1/d^{3n}$ for all $i$.
Having calculated the eigenvalues $\mu_i$, the entropy is finally
given by 
\begin{eqnarray}
	&&S_2^\epsilon(\rho_{XE})=-\log_2\left(\sum_{i=0}^{b^-}n_ix^2+
	\sum_{i=b^-+1}^{b^+-1}n_i\lambda_i^2+\right.\hspace{0.5cm}\nonumber\\
	&&\hspace{5cm}\left.\sum_{i=b^+}^mn_iy^2\right).
\end{eqnarray}

\subsection{Calculation of $H_0^\epsilon(\vec X|\vec Y)$}

Recall the definition of the conditional $\epsilon$-smooth Renyi  entropy of 
order zero,
\begin{equation}
	H_0^\epsilon(\vec X|\vec Y):=\min\limits_{\A:P(\A)\ge1-\epsilon}
	\left(\max\limits_{\vec y}\log_2|\P_{\A\vec y}|\right),
\end{equation}
where we have introduced $\P_{\A\vec y}:=\{\vec x:P_{\vec X\A|\vec Y=\vec 
y}(\vec x)>0\}$. First note that $H_0^\epsilon(\vec X|\vec Y)$ depends only on 
the number of elements in the set $\P_{\A\vec y}$, i.e. on the number of 
non-zero entries in the probability distribution $P_{\vec X\A|\vec Y=\vec y}$.  
Since in our case all values of $P_{\vec X|\vec Y=\vec y}$ are non-zero for all 
$\vec y$ (except for the case of perfect correlations, i.e. $\beta_0=1$), the 
maximization over $\vec y$ can be omitted.  Thus the only restriction on the 
number of non-zero probabilities comes from $\A$. The minimization over all 
these events occurring with probability larger or equal to $1-\epsilon$ can be 
tackled in the following way: All relevant events $\A$ need to be of the form 
$[\vec X=\vec x_1]\vee\dots\vee[\vec X=\vec x_k]$, with $\sum_{i=1}^kP_{\vec 
X}(\vec x_i)\ge1-\epsilon$. Since we are looking for the smallest set 
$\P_{\A\vec y}$ ($\vec y$ being arbitrary), we are interested in those events 
which are most restrictive, i.e. which have $k$ as small as possible. This 
means we need to find the smallest number $k$ such that the sum of the $k$ 
largest probabilities in $P_{\vec X|\vec Y=\vec y}$ is greater or equal to 
$1-\epsilon$. To this end we look at the probability 
distribution~(\ref{eq:h024hsdbaejfd234}) and find the following probabilities 
$p_l$ and occurrences $n_l$, when we condition on a certain value $\vec y$:
\begin{eqnarray}
	p_l&:=&\beta_0^l\beta_1^{n-l}\\
	n_l&:=&\genfrac(){0pt}{}{n}{l}(d-1)^{n-l}
\end{eqnarray}
In analogy to the calculation of $S_0^\epsilon(\rho_E)$, define
\begin{equation}
	s_r:=\sum\limits_{i=1}^rn_{n-i+1}p_{n-i+1},
\end{equation}
with $0\le r\le n+1$,
to be the sum of the $r$ largest \emph{different} probabilities $p_l$.
Then the smallest number $b$ such that the sum of the largest $b$
\emph{different} probabilities is greater or equal than $1-\epsilon$
is given by
\begin{equation}
	b:=\min\{r:s_r\ge1-\epsilon\}.
\end{equation}
With these definitions we find
\begin{equation}
	k=\sum\limits_{i=1}^bn_{n-i+1}-\left\lfloor\frac{s_b-(1-\epsilon)}
	{p_{n-b+1}}\right\rfloor.
\end{equation}
%or $k=d^n$ if $b=n+1$, which corresponds to the special case where
%$\epsilon=0$. 
Finally, we arrive at
\begin{equation}
	H_0^\epsilon(\vec X|\vec Y)=\log_2k.
\end{equation}

%%%%%%%%%%%%%%%%%%%%%%%%%%%%%%%%%%%%%%%%%%%%%%%%%%%%%%%%%%%%%%%%%

\section{Results}
\label{sec:k9gh3hvsbndfvf9234}

In the previous section, we calculated the entropies involved in the formula 
for the achievable key length~(\ref{eq:jkfhw38hfa0gv3w4t}). Each entropy is 
given as a simple function that can be evaluated numerically in a very 
efficient way with only $\mathcal O(n)$ running time. Note that all results are 
exact (up to machine precision), since no approximations are needed at all.  
Still, the parameter $n$ (the number of signals) is crucial in the 
implementation and we are limited to values of the order $10^4$ in this 
quantity. However, this is not a conceptual limitation: using more powerful 
computers, it is feasible to push this limit further, but we do not believe 
that this approach would yield surprising results, in view of the results 
presented in this section.
\begin{figure}[tbp]
	\begin{centering}
		\includegraphics[width=8cm]{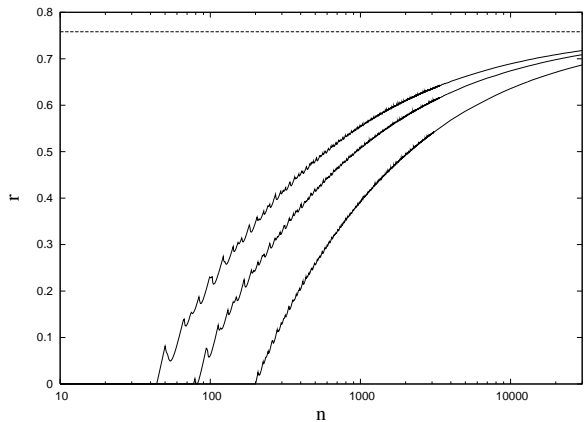}
		\caption{Key rate versus signal number $n$ for three different values 
		of the security parameter (from top to bottom: $\epsilon=0.5,0.2,0.01$) for 
		a fixed error rate in the sifted key $1-\beta_0=0.02$. The dashed line is a 
		lower bound
		of the asymptotic 
		value $\lim_{\epsilon\rightarrow0}\lim_{n\rightarrow\infty}\ell/n=
		0.758059$ found in~\cite{ren05:sec_proof_pa}.}
	\label{fig:g2h40sdhflk243fg}
	\end{centering}
\end{figure}

The scenario that we are investigating is described by the following 
parameters: The number $n$ of (quantum) signals sent from Alice to Bob which 
are kept during the sifting step, the error rate in the sifted key $1-\beta_0$ 
(see Eq.~(\ref{eq:n2492mvsssdfsasaa})), the security parameter $\epsilon$, and 
the dimensionality $d$ of the quantum systems sent from Alice to Bob. For 
better accessibility, we plot the secret key \emph{rate} $r$, which is defined 
as $r=\ell/n$, rather than the key \emph{length} $\ell$. 

Figure~\ref{fig:g2h40sdhflk243fg} shows a plot of the obtainable key rate $r$, 
as a function of the number $n$ of signals that were measured in the same basis 
by Alice and Bob. In this example we keep the error rate fixed at 
$1-\beta_0=0.02$ and show plots for different security parameters $\epsilon$.  
The error rate is chosen such that we are looking at the regime where the key 
rate is large and where a simple pre-processing does not seem to play any 
role~\cite{ren05:sec_proof_pa}. For comparison, we also plot a lower bound on 
the secret key which holds for any eavesdropping attack, but is only exact in 
the limiting case $n\rightarrow\infty$; this result was recently derived by 
Renner et. al~\cite{ren05:sec_proof_pa}. Our key rates approach the asymptotic 
value $r=0.758059$ as $n$ grows. From the plot, we recover the result found 
in~\cite{ren05:sec_proof_pa} that in the limit $n\rightarrow\infty$, the 
dependence on the security parameter $\epsilon$ becomes negligible, as the 
three curves for different $\epsilon$ approach each other. Note that for a 
small number of signals the secret key rate shows a considerable deviation from 
the asymptotic value. For a value of $n=10^4$, however, the key rate for even a 
small $\epsilon=0.01$ reaches already over 83\% of the asymptotic value. To 
give a comparison with experimental implementations, e.g. the number of signals 
$n$ (after sifting, but before classical post-processing) in the experiment 
described in~\cite{pop04:exp_ent_photons} is of the order of $10^5$.

\begin{figure}[tbp]
	\begin{centering}
		\includegraphics[width=8cm]{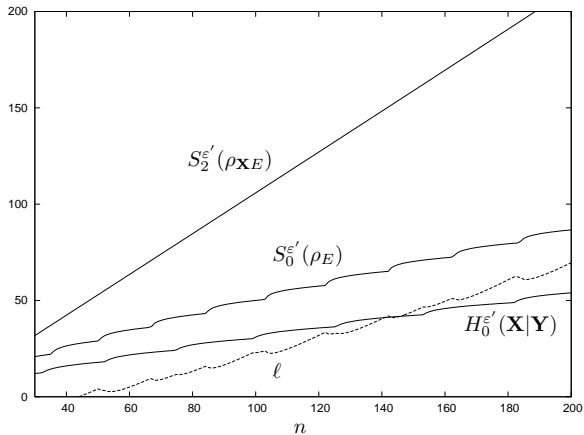}
		\caption{Plot of the three entropies in Eq.~(\ref{eq:jkfhw38hfa0gv3w4t}) 
		constituting the secret key length $\ell$ up to an additive term 
		$2\log_2(1/\epsilon)$. In this example we have $1-\beta_0=0.02$ and 
		$\epsilon=0.5$.}
	\label{fig:jg24sa4332jsknva}
	\end{centering}
\end{figure}

A prominent feature of our results are the ``oscillations'' of the achievable 
key rate, the amplitude of which decreases as $n$ increases. Analytically, the 
oscillations arise from the structure of $\ell$ given in 
Eq.~(\ref{eq:jkfhw38hfa0gv3w4t}), being the difference of the three monotonic 
functions $S_2^{\epsilon'}$, $S_0^{\epsilon'}$, and $H_0^{\epsilon'}$ where the 
last two are smoothened versions (see Fig.~\ref{fig:jg24sa4332jsknva}) of a 
non-continuous function. In the limit $n\rightarrow\infty$, the 
non-continuities disappear, leading to a monotonic key rate. Up to now, we can 
give no physical explanation for the non-monotonicity, besides the fact that 
our formula is just an achievable key rate and thus only a lower bound on the 
optimal key rate.  Moreover, we disregarded the classical pre-processing step 
in our analysis, and thus the key rate might also increase in some cases.  Note 
that up to now, no one-way pre-processing protocols except for the addition of 
noise~\cite{ren05:sec_proof_pa} have been studied. It was found that the 
addition of noise has no effect on the key rate if the correlations between 
Alice and Bob are almost perfect (as in Fig.~\ref{fig:g2h40sdhflk243fg}), but 
the rate can be increased in the region where 
$0.88\lesssim\beta_0\lesssim0.92$.  

\begin{figure}[tbp]
	\begin{centering}
		\includegraphics[width=8cm]{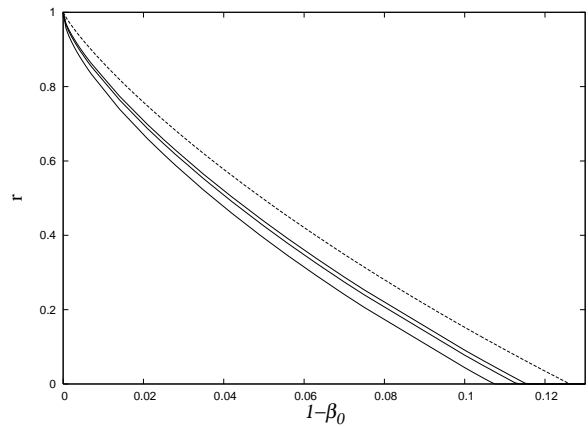}
		\caption{Key rate $r$ plotted versus the error rate in the sifted key,	
		$1-\beta_0$, for a fixed number of signals $n=20,000$. The three solid 
		lines correspond to different security parameters (from top to bottom: 
		$\epsilon=0.5,0.2,0.01$). The dashed line is again the asymptotic value 
		$\lim_{\epsilon\rightarrow0}\lim_{n\rightarrow\infty}l/n$.}
	\label{fig:lkg249gj2wbnsdmf2445}
	\end{centering}
\end{figure}

The dependence of the secret key rate on the error rate $1-\beta_0$ is 
visualized in Fig.~\ref{fig:lkg249gj2wbnsdmf2445}: The secret key rate is only 
non-negative for error rates smaller than $\approx0.11$ and gets larger as the 
error rates $1-\beta_0$ is decreased. The key rate for finite $n$ is always 
smaller than the asymptotic value (unless $1-\beta_0=0$), and it increases as 
$\epsilon$ increases, i.e. as the required security decreases.

\begin{figure}[tbp]
	\begin{centering}
		\includegraphics[width=8cm]{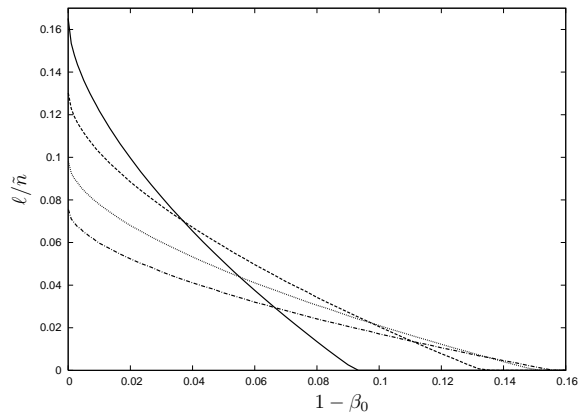}
		\caption{Effective key rate for dimension $d=2$ (solid line), $d=3$ (dashed 
		line), $d=4$ (dotted line), and $d=5$ (chain dotted line) for a fixed 
		$\tilde n=20000$ and $\epsilon=0.1$, plotted versus the error rate in the 
		sifted key $1-\beta_0$.}
	\label{fig:0349tjndg8h24fhsdg}
	\end{centering}
\end{figure}

Since our formulas are valid not only for qubits, but also for 
higher-dimensional systems, we can study the influence of the dimensionality on 
the obtainable key rate. To be able to compare the efficiency of encoding the 
information in $d=2,3,4$ dimensions, we introduce the quantity $\tilde 
n:=n'd=n(d+1)d$ which quantifies the total resources needed in the protocol: We 
have already mentioned that the number of signals before ($n'$) and after the 
sifting ($n$) are related by $n=n'/(d+1)$, where $d+1$ is the number of 
different encodings used (we consider the ``tomographic protocol'').  The 
factor $d$ accounts for the dimension of the single quantum system. We compute 
the ``effective key rate '' $\ell/\tilde n$, i.e.  the key length, measured in 
bits, divided by the ``total dimensionality'' of the Hilbert space of all 
signals of the raw key (before the sifting). In this way we have quantified the 
rate with respect to the number of initial resources needed to create the key.
Recall that $1-\beta_0=(d-1)\beta_1$ is the error rate (in the limit 
$n\rightarrow\infty$) in the sifted key, which is called quantum bit error rate 
(QBER) in the case of dimension $d=2$. This quantity gives the fraction of 
errors per $d$it in the sifted key, which makes it difficult to compare 
different dimensions, unless one can make reasonable statements about how the 
error rate $1-\beta_0$ scales with $d$, i.e. how the eavesdropper treats 
different dimensions. Keeping this problem in mind, we see in 
Fig.~\ref{fig:0349tjndg8h24fhsdg} the dependence of the effective key rate 
$\ell/\tilde n$ on the error rate $1-\beta_0$, for a fixed $\tilde n=20000$ and 
security parameter $\epsilon=0.1$. We can read off the maximal tolerable error 
rate for which a secret key can still be extracted and fortify the result found 
in~\cite{bru02:optimal_eve}, namely that the robustness of a QKD protocol 
increases as the dimension $d$ of the quantum systems increases.  This result 
also holds if sifting is disregarded, i.e.  if we keep $dn$ fixed and look at 
$\ell/(dn)$.  On the other hand, if Alice and Bob are highly correlated 
($\beta_0\rightarrow1$), we find the reverse dependence on the dimension: A 
qubit system yields the highest effective key rate and this rate decreases as 
the dimension $d$ increases.

%%%%%%%%%%%%%%%%%%%%%%%%%%%%%%%%%%%%%%%%%%%%%%%%%%%%%%%%%%%%%%%%%

\section{Conclusions}
\label{sec:fh284290uoj4jhre2}

We have developed a method for the explicit calculation of the secret key rate 
in quantum key distribution with a {\em finite} number of signals $n$,  under 
the assumption that the eavesdropper only conducts symmetric collective 
attacks, i.e. the state shared by Alice and Bob after the quantum part
of the protocol (cf.  section~\ref{sec:h30gfhwsefoj234twef}) has the form 
$\rho_{AB}^{\otimes n}=
[(\beta_0-\beta_1)\proj{\phi_d^+}+\beta_1\1/d]^{\otimes n}$. At this
step, Alice and Bob have to measure this state in the computational
basis to obtain the classical bit strings that are the starting point
of the classical post-processing. This means that \emph{any} protocol
in which Alice and Bob can ensure that they share such a state and
which uses privacy amplification is covered by our analysis. In reality, 
obtaining knowledge about $\rho_{AB}^{\otimes n}$ is a hard task, but
we believe that our analysis of the idealized case helps in
solving the challenge of a finite key analysis of a more general scenario.

We have shown that the secret key rate obtainable by our protocol strongly 
depends on the number of quantum signals sent. Our results suggest that for 
signal numbers larger than $n\sim10^4$, the asymptotic value for the key rate 
found by~\cite{ren05:sec_proof_pa} is a good approximation. However, for 
smaller values of $n$, we find a significantly lower value. This is remarkable 
in particular because we restricted our analysis to a symmetric eavesdropping 
strategy, thereby weakening Eve's power and potentially increasing the 
obtainable key rate. In contrast, the result found in~\cite{ren05:sec_proof_pa} 
covers \emph{all} eavesdropping attacks and thus the asymptotic value of $r$ is 
already based on pessimistic assumptions. Therefore, our results suggest that
for scenarios with only a few number of signals, significant deviations of the 
key rate from the asymptotic value are to be expected.

A popular task in the analysis of quantum key distribution is the 
characterization of the \emph{threshold QBER}, which is the maximal quantum bit 
error rate, for which the protocol still yields a non-vanishing key rate.  
However, even a high threshold QBER does not guarantee a feasible protocol, as 
the key rate might be arbitrarily close to zero or increase very slowly with
decreasing QBER. Our results on the other hand quantitatively characterize the 
secret key rate with respect to all parameters of the protocol. In particular, 
we have shown that for $d$-dimensional generalizations of the six-state 
protocol, larger dimensions give a higher robustness, i.e.  more noise is 
tolerable, but smaller dimensions yield a higher key rate if the the 
correlations between Alice and Bob are already high.

%%%%%%%%%%%%%%%%%%%%%%%%%%%%%%%%%%%%%%%%%%%%%%%%%%%%%%%%%%%%%%%%%

\section{Acknowledgements}

We would like to thank Barbara Kraus, Norbert L\"utkenhaus, and in particular 
Renato Renner for valuable discussions. This work was supported by the European 
Commission (Integrated Project SECOQC).

%%%%%%%%%%%%%%%%%%%%%%%%%%%%%%%%%%%%%%%%%%%%%%%%%%%%%%%%%%%%%%%%%

\bibliographystyle{prsty}
\bibliography{bla}

\end{document}